\documentclass[10pt]{IEEEtran}
\usepackage{stfloats}

\usepackage{enumerate}
\usepackage{algorithm,algpseudocode}
\usepackage{setspace,amsmath,latexsym,cite,amssymb,epsfig,amsfonts}
\usepackage{url,cite}
\usepackage{graphicx}
\usepackage{psfrag}
\usepackage{footmisc}
\usepackage{multirow}
\usepackage{color}
\usepackage{multicol}
\usepackage{mathtools}
\usepackage{subcaption}
\usepackage{graphicx}
\usepackage{amssymb}
\usepackage{amsmath}
\usepackage{epstopdf}
\usepackage{geometry}
\usepackage{float}
\usepackage{balance}

\twocolumn

\allowdisplaybreaks[4]

\title{Hybrid Generative Semantic and Bit Communications in Satellite Networks: Trade-offs\\in Latency, Generation Quality, and Computation}

\author{\IEEEauthorblockN{Chong Huang$^1$, Gaojie Chen$^2$, Jing Zhu$^2$, Qu Luo$^1$, Pei Xiao$^1$, Wei Huang$^2$, Rahim Tafazolli$^1$} \\
\IEEEauthorblockA{
$^1$5GIC \& 6GIC, Institute for Communication Systems (ICS), University of Surrey, GU2 7XH, UK\\
Email: $\{$chong.huang, q.u.luo, p.xiao, r.tafazolli$\}$@surrey.ac.uk \\
$^2$School of Flexible Electronics, Sun Yat-sen University, 510220, China\\
Email: gaojie.chen@ieee.org, zhuj229@mail.sysu.edu.cn, huangw323@mail.sysu.edu.cn \\
}
}

\begin{document}
\captionsetup[figure]{name={Fig.},labelsep=period}

\begin{singlespace}
\maketitle
\end{singlespace}

\thispagestyle{empty}
\pagestyle{empty}
\begin{abstract}
As satellite communications play an increasingly important role in future wireless networks, the issue of limited link budget in satellite systems has attracted significant attention in current research. Although semantic communications emerge as a promising solution to address these constraints, it introduces the challenge of increased computational resource consumption in wireless communications. To address these challenges, we propose a multi-layer hybrid bit and generative semantic communication framework which can adapt to the dynamic satellite communication networks. Furthermore, to balance the semantic communication efficiency and performance in satellite-to-ground transmissions, we introduce a novel semantic communication efficiency metric (SEM) that evaluates the trade-offs among latency, computational consumption, and semantic reconstruction quality in the proposed framework. Moreover, we utilize a novel deep reinforcement learning (DRL) algorithm group relative policy optimization (GRPO) to optimize the resource allocation in the proposed network. Simulation results demonstrate the flexibility of our proposed transmission framework and the effectiveness of the proposed metric SEM, illustrate the relationships among various semantic communication metrics.
\end{abstract}

\section{Introduction}
\IEEEPARstart{I}{n} recent years, satellite communications have emerged as a significant research direction for future sixth-generation (6G) networks due to their capability to provide extensive coverage and reliable connectivity for global users, especially in remote areas such as mountainous regions and oceans, satellite communications effectively overcome geographical constraints compared to traditional terrestrial communication networks \cite{9508471}. Furthermore, satellite communications can offer stable communication links during special scenarios such as natural disasters and emergency rescue. Recently, numerous emerging satellite network projects such as Starlink, have developed rapidly in wireless communications. Moreover, the potential applications of satellite communication in areas such as edge computing, the Internet of Things (IoT), and cognitive communications are gradually being realized, making it a critical method to achieve seamless connectivity \cite{10440193}.

However, the link budget of satellite communications is constrained by many limitations such as transmission power, bandwidth constraints, and scarce satellite orbital resources \cite{9295418}. Therefore, it is necessary for future satellite communications to explore innovative wireless transmission technologies that effectively reduce the size of transmitted data and improve transmission efficiency. Traditional bit communication methods focus on the entire data transmission, usually overlook the varying importance of data in different scenarios. Recently, semantic communications provide an innovative solution to address this issue \cite{10183794}. Semantic communications extract the information features from data by leveraging intelligent information processing and semantic understanding, it significantly reduces bandwidth requirements for data transmissions and greatly improves the transmission efficiency \cite{10639525}. Such a semantic-oriented communication method is especially crucial in satellite communication environments, which typically exhibit low signal-to-noise ratios (SNR) \cite{9252948}. It can effectively alleviate link budget constraints and substantially enhance the overall performance and reliability of future communication systems. In \cite{10445211}, semantic communications were utilized to enhance the offloading efficiency for edge computing in satellite networks. To enhance the transmission efficiency and minimize the total transmission cost in space-air-ground integrated networks (SAGINs), an alternating algorithm was introduced to optimize the resource allocation for semantic communications in \cite{10901382}.

Although semantic communications can effectively reduce the transmission bandwidth requirements, it requires high computational resources in compression and generation process \cite{10387520}. Thus, its communication efficiency and performance depend on the balance among reconstruction quality at the receiver, transmission delay, and computational consumption. In satellite communications, due to limited transmit power, high latency caused by long-distance transmission, and constrained computational resources at both the transmitter and the receiver, it becomes very important to ensure accurate and timely reconstruction of semantic information under limited computational resource. The study in \cite{10909615} investigated the advantage of leveraging computational consumption in semantic communications to reduce communication load in wireless networks. In \cite{huang2024deep}, a learning-based optimization was utilized to achieve the trade-off between delay and generation quality in satellite communications.

To investigate a novel semantic communication framework which can adapt to dynamic satellite-ground communication environments while considering trade-offs among various factors influencing communication efficiency, we propose a multi-layer hybrid bit and generative semantic communication framework, and a novel semantic metric for balancing latency, computational consumption, and reconstruction quality. The main contributions of this paper are summarized as follows.
\begin{itemize}
  \item We consider a multi-layer hybrid bit and generative semantic communication framework in satellite communications. Our framework considers both traditional bit transmission mode and multi-layer semantic transmission modes with varying compression ratios and computational costs. Moreover, inter-satellite links (ISLs) are considered in satellite networks to forward data beyond satellite communication coverage.

  \item We propose a novel semantic efficiency metric (SEM) to evaluate trade-offs among transmission latency, computational consumption, and semantic reconstruction quality in satellite networks. Furthermore, we utilize a deep reinforcement learning (DRL) method to optimize resource allocation to maximize SEM in satellite cmmuunications.

  \item Simulations validate the proposed communication framework and optimization algorithm, and demonstrate the effectiveness of the proposed metric in dynamic satellite networks.
\end{itemize}

The rest of this paper is organized as follows: Section \ref{se:sm} introduces the system model, the hybrid transmission framework and the optimization problem. In Section \ref{sec:DRL}, the proposed DRL-based optimization algorithm is presented. \ref{sec:sim} verifies the proposed framework and metric. Finally, Section \ref{sec:con} concludes this work.

\section{System Model and Problem Formulation} \label{se:sm}
\subsection{System Model}
\begin{figure}[t!]
  \centering
  \centerline{\includegraphics[width=0.47\textwidth]{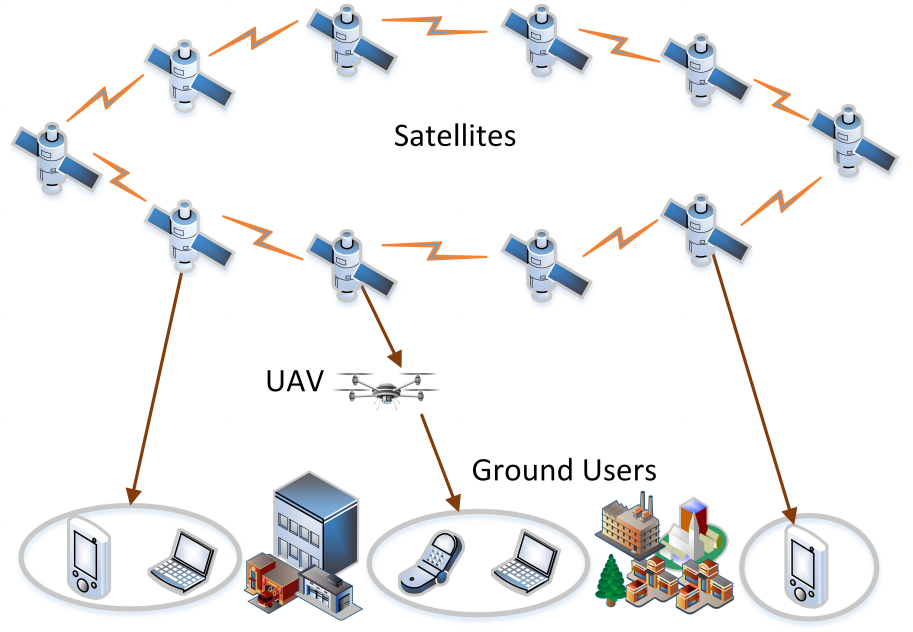}}
 \caption{System model of the proposed satellite network.} \label{fig:SM}
\end{figure}
As shown in Fig. \ref{fig:SM}, we consider a satellite network that includes $M$ low Earth orbit (LEO) satellites $S_m$ ($m \in \mathcal M = \{1, 2, ..., M\}$), $N$ UAVs $U_n$ ($n \in \mathcal N = \{1, 2, ..., N\}$), and $L$ ground users $G_l$ ($l \in \mathcal L = \{1, 2, ..., L\}$). LEO satellites need to transmit data to ground users based on transmission tasks. Since reconstructing data through semantic communication requires significant computational resources, some ground users may face computational resource limitation or inability to directly receive satellite signals. Therefore, UAVs are employed to replace these constrained ground users by receiving and reconstructing semantic information before forwarding the generated data to the corresponding users. The location of $U_n$ at time slot $t$ is $q_n(t) \in {\mathcal Q}$, where $q_n(t)$ is the 3D coordinates with height $H_q$, one UAV serves no more than one ground user at a given time slot, the channel coefficient between ground user $G_l$ and UAV $U_n$ follows Rician fading as
\begin{equation}\small\label{eqh1}
h_{l,n}= \sqrt{\frac{\kappa}{\kappa+1}} \frac{\bar{H}_{l,n}}{d_{l,n}^{-{\alpha_L}/2}}+\sqrt{\frac{1}{\kappa+1}} \frac{\hat{H}_{l,n}}{d_{l,n}^{-{\alpha_N}/2}} ,
\end{equation}
where $\kappa$ denotes the Rician factor, $\bar{H}_{l,n}$ and $\hat{H}_{l,n}$ denotes the line-of-sight (LoS) and non-line-of-sight (NLoS) coefficients, respectively, $|\bar{H}_{l,n}| = 1$, $\hat{H}_{l,n}$ follows zero-mean unit-variance Gaussian distribution, $d_{l,n}$ is the distance between ground user $G_l$ and UAV $U_n$, $\alpha_L$ and $\alpha_N$ represents the path loss exponents of LoS and NLoS, respectively. The channel capacity of link between $G_l$ and $U_n$ is expressed as
\begin{equation}\small\label{eqc1}
R_{l,n} = B_{l,n} {\rm{log_{2}}} \left( 1 + \frac{P_{U_n} |h_{l,n}|^2} {{{\sigma}_{l}^2} } \right),
\end{equation}
where $B_{l,n}$ denotes the bandwidth for the channel between ground user $G_l$ and UAV $U_n$, $P_{U_n}$ presents the transmit power of UAV $U_n$, ${{\sigma}_{l}^2}$ is the additive white Gaussian noise (AWGN). We assume $o_{m,l,n} \in \{0, 1\}$ is the indicator for satellite-UAV-user transmissions, where $o_{m,l,n} = 1$ denotes UAV $U_n$ is used for the transmission between LEO $S_m$ and ground user $G_l$, otherwise $o_{m,l,n} = 0$.

For satellite to ground user transmissions, we formulate the channel as $h_{m,l} = \frac{\sqrt{\beta_m}\lambda}{4 \pi d_{m,l}} e^{j {\gamma_m}}$, where $\beta_m$ is the antenna gain of satellite $S_m$, $\lambda$ denotes the wavelength, $d_{m,l}$ indicates the distance between LEO and ground user, and $\gamma_m$ is the phase component of the LEO satellite. Due to the long distance transmission and the high-speed mobility of LEOs, we consider a outdated channel state information (CSI) model as \cite{9535285} $bar{h}_{m,l} = \xi h_{m,l} + \sqrt{1 - \xi^2} \hat{h}_{m,l}$, where $\xi = \bar{\varpi} (2 \pi {\bar{\epsilon}}_{m,l} T_{m,l})$, with $\bar{\varpi}$ denotes the Bessel function of 0 order, ${\bar{\epsilon}}_{m,l}$ indicates the maximum Doppler frequency, and $T_{m,l}$ is the delay of transmissions between LEO and ground user. $\hat{h}_{m,l}$ follows the Gaussian distribution with variance $h_{m,l}$. In our system, we need to optimize the transmission resource based on the outdated CSI while the real channel capacity is based on the exact channel coefficient as
\begin{equation}\small\label{eqc2}
R_{m,l} = B_{m,l} {\rm{log_{2}}} \left( 1 + \frac{P_{S_m} |h_{m,l}|^2}{{{\sigma}_{l}^2}} \right),
\end{equation}
where $B_{m,l}$ denotes the corresponding bandwidth, $P_{S_m}$ is the transmit power of LEO $S_m$. Similarly, we can obtain the channel coefficient $h_{m,n}$, outdated channel coefficient $\bar{h}_{m,n}$ and channel capacity $R_{m,n}$ for transmissions between LEO $S_m$ and UAV $U_n$ as described above.

Moreover, due to the limited coverage area of LEO satellites, if a LEO cannot directly transmit task data to the corresponding ground user, it can send the data to another LEO satellite which can access to the corresponding ground user to complete the transmission. The LEO coverage model is the same as in \cite{10440193}, the minimum elevation angle of LEOs is $\zeta$. The channel capacity for a ISL between LEO $S_{m1}$ and $S_{m2}$ is \cite{10440193}
\begin{equation}\small\label{eqISL}
R_{m_1,m_2} = B_{m_1,m_2} {\rm{log_{2}}} \left( 1 + \frac{P_{S_{m_1}} |\delta_{\rm max}|^2}{\varepsilon \rho B_{{m_1},{m_2}} \left(\frac{{4 \pi d_{m_1,m_2} f_{\rm S}}}{c}\right)^2} \right),
\end{equation}
where $m_1 \neq m_2, m_1 \in \mathcal M, m_2 \in \mathcal M$, $B_{m_1,m_2}$ denotes the bandwidth for this ISL, $P_{S_{m_1}}$ indicates the transmit power of LEO $S_{m_1}$, $\delta_{\rm max}$ is the peak antenna gain of $S_{m_1}$, $\varepsilon$ denotes the Boltzmann constant, $\rho$ presents the thermal noise, $d_{m_1,m_2}$ is the distance between two LEOs, $f_{\rm S}$ presents the carrier frequency, $c$ denotes the speed of light. We assume $z_{m_1, m_2} \in \mathcal{Z}$ as the indicator for ISLs, where $z_{m_1, m_2} = 1$ denotes LEO $S_{m_1}$ send task data to $S_{m_2}$ via ISLs, otherwise $z_{m_1, m_2} = 0$. Each LEO can only send data via ISL once at a given time slot.

\subsection{Hybrid Bit and Semantic Transmission Model}
\begin{figure*}[t!]
    \centering
    \includegraphics[width=0.85\linewidth]{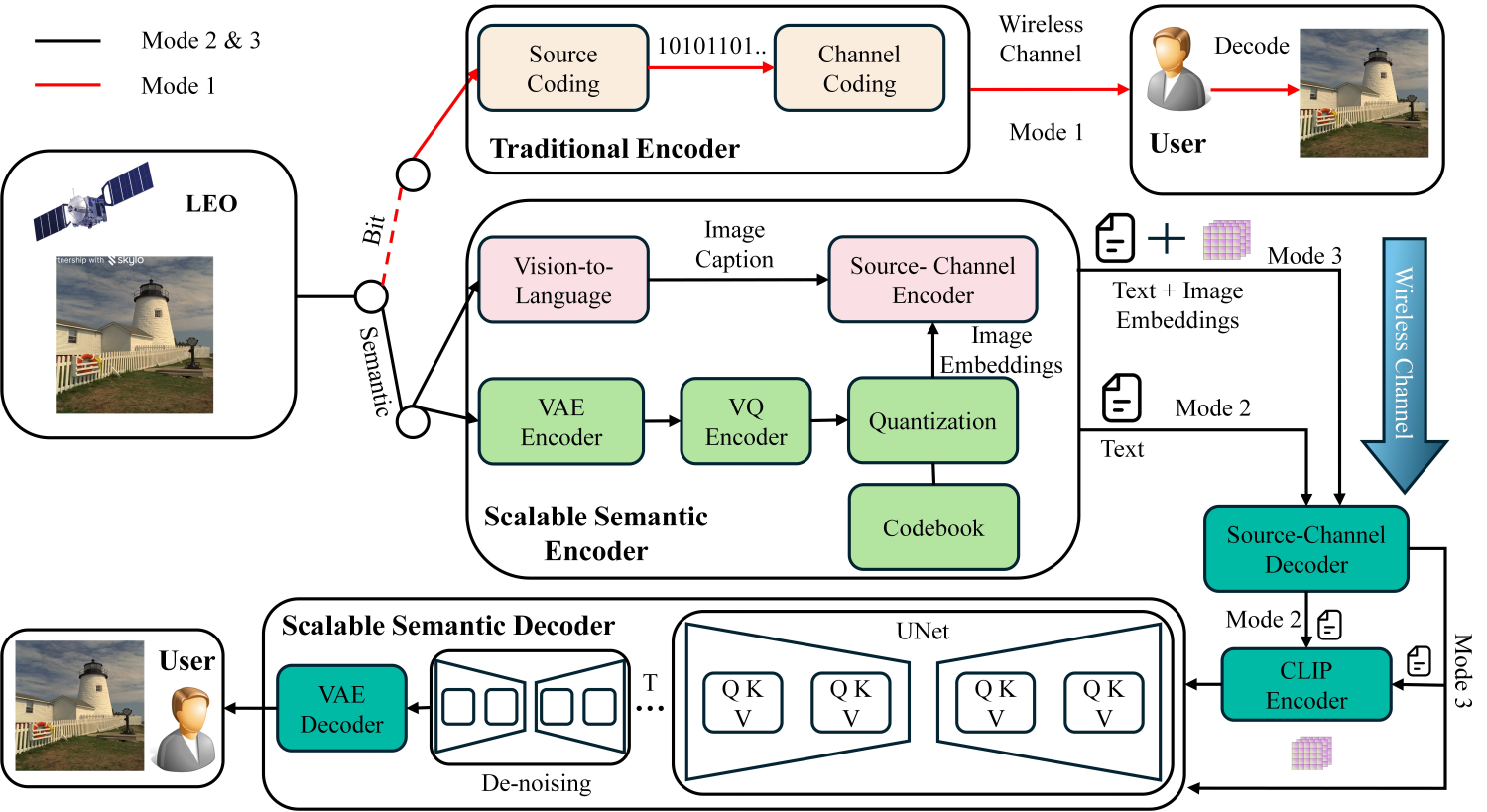}
    \caption{The hybrid bit and semantic communication framework.}
    \label{sc_framwork}
\end{figure*}
Considering the dynamic environment of satellite communications and computational limitations of users, we design a multi-layer hybrid transmission framework consisting of three modes: traditional bit transmission mode, text transmission mode, and combined text-image feature transmission mode. Moreover, the combined text-image feature transmission mode has multiple layers to further adapt to varying SNR conditions and task requirements. The details of three modes are as follows:

\begin{enumerate}
    \item \textbf{Mode 1}: Traditional bit transmission without semantic extraction and compression. The original image is compressed by using conventional algorithms such as JPEG 2000 and then transmitted to the receiver. This mode requires minimal computational resources and provides very high quality reconstructed images, but demands high transmission SNR.
    \item \textbf{Mode 2}: Semantic extraction of original image information to generate text, which are transmitted to the receiver for reconstructing the image. This mode requires low SNR and high computational resources. However, the quality of the reconstructed images may be insufficient.
    \item \textbf{Mode 3}: Semantic extraction of both text and image features from the original image for transmission. This mode demands high computational resources and moderate transmission SNR, and achieves high quality of reconstructed images.
\end{enumerate}

For the proposed semantic communication modes as shown in Fig. \ref{sc_framwork}, we utilize the BLIP-2 algorithm at the transmitter to extract information from the original image and convert it to text, and further transfer the text to text features by using the contrastive language-image pre-training (CLIP) method. Moreover, we use variational autoencoder (VAE) and vector quantization encoder (VQE) to extract image features, and select an appropriate codebook to output corresponding image embeddings. At the receiver, we reconstruct the image based on the received text and image embeddings through an $e$ steps de-noising process. For Mode 2, reconstruction is performed based on text only, while Mode 3 utilizes both text and image features for reconstruction. Thus, we can balance latency, computational cost, and reconstruction quality not only by selecting transmission modes but also by varying the number of de-noising steps. Notice that there are several approaches for adjusting the relationship among metrics in semantic communications, this work focuses on studying the trade-offs among these metrics rather than exploring novel semantic communication algorithms.

To investigate the trade-offs among computational consumption, latency, and reconstruction quality in semantic communications, we propose a novel semantic communication efficiency metric SEM which is defined as
\begin{equation}\small\label{metric}
SEM_k = \theta_{D} (D_{k, \rm max} - D_k) + \theta_{R} (\vartheta_k - \vartheta_{k, \rm min}) - \theta_{C} F_k,
\end{equation}
where $k \in {\mathcal K}$ denotes $k$th transmission task, we assume there are $K$ transmission tasks in the proposed network. $\theta_{D}$, $\theta_{R}$ and $\theta_{C}$ are the weights for latency, reconstruction quality, and computational consumption, respectively, $\theta_{D} + \theta_{R} + \theta_{C} = 1$. $D_{\rm max}$ and $\vartheta_{\rm min}$ indicate the maximum delay threshold and minimum quality threshold for $k$th task, respectively. $D_k$, $\vartheta_k$ and $F_k$ are the delay, reconstruction quality, and computational cost for $k$th task, respectively. Notice that we have normalized all parameters in SEM to the range (0, 1) for clarity. From \eqref{metric} it can be observed that both delay and computational consumption negatively influence SEM, while reconstruction quality positively contributes to this metric.

\subsection{Problem Formulation}
In the satellite network, we need to optimize the transmission mode selection, UAV trajectory, UAV selection, ISL selection, and de-noising steps to maximize the average semantic communication metric SEM. The problem formulation is as
\begin{align}
    &\max_{\mathcal{W}(t), {\mathcal Q}(t), \mathcal{O}(t), \mathcal{Z}(t), \mathcal{E}(t)} \frac{1}{K} \sum_{k=1}^{K} SEM_k,\label{SecrecyFunc}\\
    {\rm s.t.}&~ \sum_{l=1}^{L}\sum_{n=1}^{N} o_{m,l,n} \in \{0, 1\}, \forall m \in {\mathcal M} \tag{\ref{SecrecyFunc}{a}}, \label{SecrecyFuncSuba}\\
    &\sum_{m=1}^{M}\sum_{l=1}^{L} o_{m,l,n} \in \{0, 1\}, \forall n \in {\mathcal N} \tag{\ref{SecrecyFunc}{b}}, \label{SecrecyFuncSubb}\\
    &\sum_{m=1}^{M}\sum_{n=1}^{N} o_{m,l,n} \in \{0, 1\}, \forall l \in {\mathcal L} \tag{\ref{SecrecyFunc}{c}}, \label{SecrecyFuncSubc}\\
    &\sum_{m_1=1}^{M} z_{m_1, m_2} \in \{0, 1\}, \forall m_2 \in {\mathcal N}, m_1 \neq m_2 \tag{\ref{SecrecyFunc}{d}}, \label{SecrecyFuncSubd}\\
    &\sum_{m_2=1}^{M} z_{m_1, m_2} \in \{0, 1\}, \forall m_1 \in {\mathcal N}, m_1 \neq m_2 \tag{\ref{SecrecyFunc}{e}}, \label{SecrecyFuncSube}\\
    &v_n \leq v_{\max}, \forall n \in {\mathcal N} \tag{\ref{SecrecyFunc}{f}}, \label{SecrecyFuncSubf}\\
    &D_k \leq D_{k,\rm max}, \forall k \in {\mathcal K} \tag{\ref{SecrecyFunc}{g}}, \label{SecrecyFuncSubg}\\
    &\vartheta_k \geq \vartheta_{k,\rm min}, \forall k \in {\mathcal K} \tag{\ref{SecrecyFunc}{h}}, \label{SecrecyFuncSubh}
\end{align}
where $\mathcal{W} = \{w_1, w_2, ..., w_K\}$ is the transmission mode selection set for tasks, $\mathcal {Q}(t) = \{q_1(t),..., q_N(t)\}$ denotes the trajectory set for UAVs at time slot $t$, $\mathcal{O}(t) = \{o_{m,l,n}, n \in {\mathcal N}, l \in {\mathcal L}, M \in {\mathcal M}\}$ is the UAV selection set, $\mathcal{Z}(t) = \{z_{m_1,m_2}, m_1 \& m_2 \in {\mathcal N}, m_1 \neq m_2\}$ presents the ISL selection, $\mathcal{E} = \{e_1, ..., e_K\}$ denotes the de-noising step set for all tasks. \eqref{SecrecyFuncSuba}-\eqref{SecrecyFuncSubc} denotes each satellite selects at most one UAV and ground user for the transmission task at a given time slot, and one ground user can only receive data from one LEO at a give time slot. \eqref{SecrecyFuncSubd} and \eqref{SecrecyFuncSube} denotes one LEO can join only one ISL transmission at a give time slot. \eqref{SecrecyFuncSubf} is the speed limitation for UAVs, where $v_n$ is the UAV speed, \eqref{SecrecyFuncSubg} and \eqref{SecrecyFuncSubh} present the latency and quality threshold for transmission tasks.

\section{DRL-Based Optimization}\label{sec:DRL}
To address the non-convex problem in \eqref{SecrecyFunc}, we utilize a DRL algorithm to maximize the metric SEM in satellite networks. First, we define the MDP elements for DRL. State $u(t) = \{t, h_{n,l}(t)_{n \in \mathcal N, l \in \mathcal L}, {\mathcal Q}(t), {D_k(t)}_{k \in \mathcal K}, {\vartheta_K(t)}_{k \in \mathcal K}\, \mu_k\}$, where $\mu_k = \{0, 1\}$ is the completion indicator for $k$th task, $\mu_k = 1$ denotes this task is completed, otherwise $\mu_k = 0$. Action $a(t) = \{\mathcal{W}(t), {\mathcal Q}(t), \mathcal{O}(t), \mathcal{Z}(t), \mathcal{E}(t)\}$ is based on the optimization variables from \eqref{SecrecyFunc}. Reward $r(t) = \frac{1}{K}\sum_{k=1}^{K} SEM_k(t)$ can help the agent in DRL to maximize the sum SEM in satellite communications. We utilize a novel DRL algorithm group relative policy optimization (GRPO), which was used to train DeepSeek, to learn the solution for problem in \eqref{SecrecyFunc}. Considering it is a variant of proximal policy optimization (PPO) and PPO has already been one of the most popular DRL algorithm, so we briefly introduce some important changes in GRPO compared to PPO, other DRL information can be found in PPO \cite{schulman2017proximal} and GRPO \cite{shao2024deepseekmath} works.

In PPO, the value function and the policy function are typically learned simultaneously. In contrast, GRPO removes value estimation by employing the group-based average reward as a baseline to evaluate the policy network. This change significantly reduces the training cost of DRL and provides a more accurate estimation for actions. The loss function in GRPO is formulated as
\begin{equation}\small\label{eq:GRPO}
\begin{split}
    Loss(\phi) = &~\frac{1}{\upsilon}\sum_{i=1}^\upsilon \bigg\{ \min \bigg[ \frac{\pi_{\phi}(a(t) | s(t))}{\pi_{\phi_{\text{old}}}(a(t) | s(t)} \hat{A}_{t}, \\
    &\text{clip} \bigg( \frac{\pi_{\phi}(a(t) | s(t))}{\pi_{\phi_{\text{old}}}((a(t) | s(t))}, 1 - \psi, 1 + \psi \bigg)  \hat{A}_{t} \bigg]\bigg\},
\end{split}
\end{equation}
where $\phi$ is the current policy network, $\phi_{old}$ denotes the old policy network, $v$ is the number of samples used for reward feedback, $\psi$ is the clip factor, $\hat{A}$ is the advantage from 4.1.2 in \cite{shao2024deepseekmath}, $\varrho$ is a parameter in GRPO. Notice that we remove the Kullback–Leibler (KL) divergence from the loss function in GRPO, as some studies suggest that removing KL divergence can further enhance the performance in GRPO. GRPO presents significant potential for using reward in DRL and smoothing out variations in rewards caused by probabilistic factors in reinforcement learning environments. Moreover, by directly using rewards rather than introducing advantage estimation through training additional value networks, GRPO significantly reduces the computational cost in training process.

\section{SIMULATION RESULTS}\label{sec:sim}
Simulation parameters are set as follows: the number of LEOs $M = 5$, the number of ground users $L = 5$, the number of UAVs $N = 2$. 1 or 2 ground users are randomly selected as which need UAV to forward data from LEOs. The altitude of LEOs is 750 km, the altitude of UAV is 100 m, the distance between LEOs are randomly generated from 200 km to 600 km, the speed of LEOs is 7.8 km/s, the speed limitation of UAVs is 12 m/s, the antenna gain of LEOs is 42 dB, the Doppler frequency is set in Ka-band, the carrier frequency of ISLs is 25 GHz, the transmit powers of LEOs and UAVs are 1 W and 0.35 W, respectively. Thermal noise is 354.81 K, the noise level is -130 dBm, $\zeta = 40 ^{\circ}$, the bandwidth of channels is 20 MHz, $\alpha_L = 2.2$ and $\alpha_N = 2.5$. Transmission tasks arrives based on FTP Model 3 \cite{3GPPFTPmodel3}, the size of original data is 3.5 MB, $K = 100$, the delay constraints are randomly distributed between 3 s and 10 s, the quality constraints measured in PSNR are randomly distributed between 12 and 22, the computational cost for one de-noise step is around 686 GFlops at the receiver, and 810 GFlops at the transmitter, the computational cost of JPEG 2000 is very small and can be neglected. GRPO parameter values are the same as in \cite{shao2024deepseekmath}. We use PPO as the benchmark in our simulations.

\begin{figure}[t!]
  \centering
  \centerline{\includegraphics[scale=0.55]{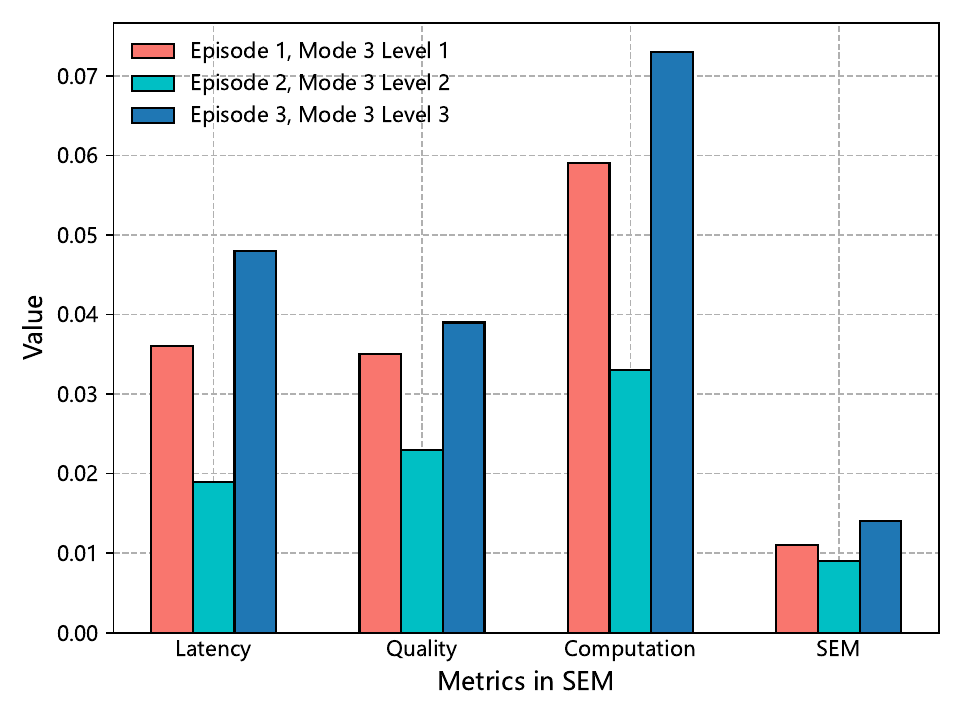}}
 \caption{\small The values of metrics in SCM versus different semantic transmission mode.} \label{fig:R0}
\end{figure}
We compare the values of latency, generation quality and computation cost in SEM in Fig. \ref{fig:R0}. `Episode 1, Mode 3 Level 1' denotes using Mode 3 to transmit a task with moderate computational resource, `Episode 1, Mode 3 Level 2' utilizes less computational resource than that in Level 1, while `Episode 1, Mode 3 Level 3' uses the highest computational consumption for compression and reconstruction. The value of each metric corresponds to the absolute value of the respective term in \eqref{metric}. It can be observed that when we select a transmission mode with low computational cost, the latency increases due to reduced compression and reconstruction process, this leads to the decrease of corresponding metric value. Similarly, the reconstruction quality declines while the negative impact of computational consumption also diminishes accordingly. With evenly distributed weights, the overall SEM value decreases in this case. Moreover, increasing computational resources enables further compression of semantic information to reduce transmission delay and potentially improve reconstruction quality. Therefore, variations in each metric significantly impact SEM value, and the weights assigned to these metrics should be carefully balanced according to practical requirements.

\begin{figure}[t!]
  \centering
  \centerline{\includegraphics[scale=0.55]{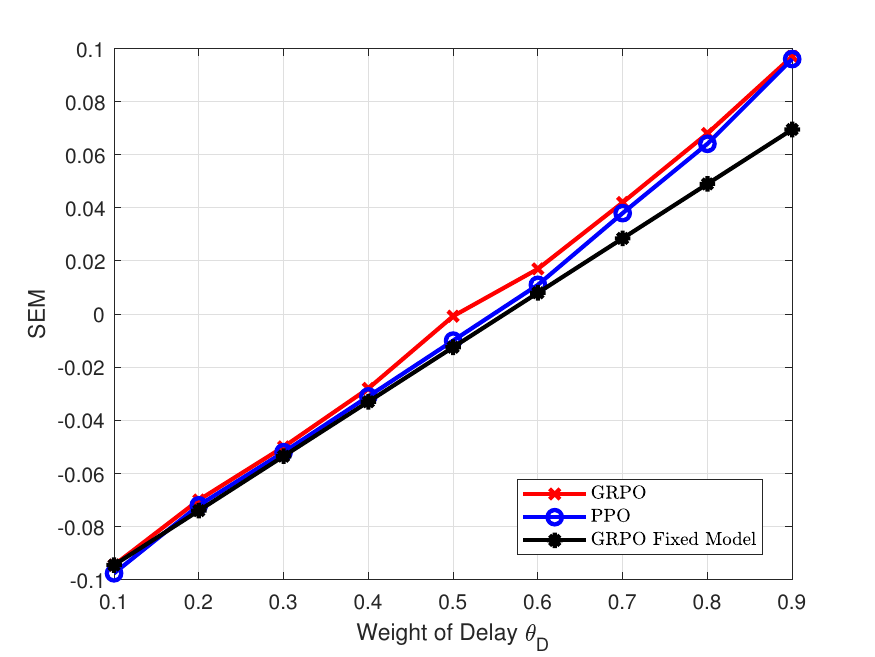}}
 \caption{\small Average SEM versus the weight of delay.} \label{fig:R1}
\end{figure}
From Fig. \ref{fig:R1}, we compare the average SEM with different value of weight for delay in SEM, where the other two weights are set equal. `GRPO Fixed Model' denotes that only use model trained for $\theta_D = 0.1$. As illustrated in the figure, SEM value increases linearly with increasing weights without weight optimization, which aligns with the theoretical insight from \eqref{metric}. However, as the weight of delay increases, resource optimization can increasingly focus on reducing transmission latency, i.e., selecting transmission modes with high compression ratios to lower the task latency and further enhance SEM. After optimizing based on the dynamic weights, it can be observed that SEM increases more than `GRPO Fixed Model', this is because that transmission mode selection and other optimization variables significantly impact delay. Moreover, the proposed algorithm outperforms PPO, this result shows its superiority in optimizing the hybrid transmission framework and analyzing the trade-off between metrics.

\begin{figure}[t!]
  \centering
  \centerline{\includegraphics[scale=0.55]{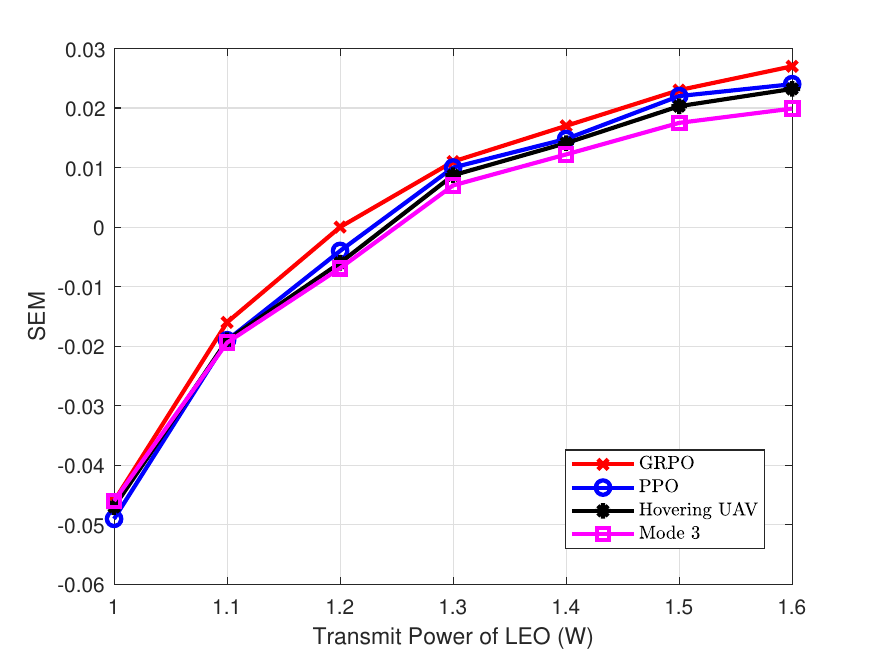}}
 \caption{\small Average SEM versus the transmit power of LEOs.} \label{fig:R2}
\end{figure}
In Fig. \ref{fig:R2}, we compare the average SEM under different LEO transmit power, with the three weights equally distributed. We introduce `Hovering UAV' and `Mode 3' as baselines for comparison, where `Hovering UAV' represents the scenario that UAVs keep in hovering state, and `Mode 3' denotes only Mode 3 is available in the hybrid transmission framework, without bit transmission and Mode 2. As can be observed from the figure, removing either trajectory optimization or the two transmission modes leads to performance degradation. This is because trajectory optimization enables the UAV to better communicate with ground users, whereas bit transmission and Mode 2 can achieve reduced latency and higher reconstruction quality under high SNR conditions. These results demonstrate the significance of each optimization variable within the proposed framework and highlight the huge optimization potential in highly dynamic satellite networks.

\section{Conclusion}\label{sec:con}
In this paper, we investigated a hybrid bit and generative semantic communication framework for satellite networks, and proposed a novel semantic efficiency metric SEM which balances latency, computational consumption, and reconstruction quality. Furthermore, we employed a multi-layer semantic transmission mode and a multi-step denoising strategy to adjust the semantic communication performance. To optimize the proposed transmission framework, we introduced a novel algorithm GRPO and confirmed its superiority in evaluating trade-offs among semantic communication metrics. As semantic communications save spectrum resources at the cost of increasing computational consumption and data quality degradation, the trade-offs among these metrics will play a key role in future wireless communication systems, especially for satellite networks where link budgets are severely constrained. In our future work, we plan to further explore the trade-offs among semantic communication metrics and provide further analysis on the weighting of metrics within SEM.


\bibliographystyle{ieeetr}
\bibliography{ref}


\end{document}